\documentclass[a4paper,12pt]{article}
\usepackage[utf8]{inputenc}
\usepackage[english]{babel}
\usepackage{amsmath, amssymb, graphicx,bm}
\usepackage{braket}
\usepackage{caption}
\usepackage{subcaption}
\usepackage{ulem}
\usepackage{multicol}

\pdfoutput=1
\usepackage{jheppubm}
\newcommand{\be}{\begin{equation}}
\newcommand{\ee}{\end{equation}}
\newcommand{\bea}{\begin{eqnarray}}
\newcommand{\eea}{\end{eqnarray}}
\newcommand{\nn}{\nonumber}

\definecolor{darkraspberry}{rgb}{0.53,0.15,0.34}

\definecolor{darkblue}{rgb}{0,0,1}

\definecolor{dgreen}{rgb}{0,0.6,0}

\definecolor{brown}{rgb}{0.59,0.29,0}

\numberwithin{equation}{section}

\begin{document}

\title{Quantum vs. Symplectic Computers}

\author{Igor Volovich}

\affiliation{Steklov Mathematical Institute, Russian Academy of Science}

\emailAdd{volovich@mi-ras.ru}

\date{June 2024}

\abstract{In this paper, we propose the concept of symplectic computers, which have the potential to be more powerful than quantum computers. Unlike quantum computing, which consists of a sequence of unitary transformations (gates) and projectors (measurements), symplectic computation involves a sequence of symplectic transformations and measurements.

The proposal to explore symplectic computers is based on the following quantum-symplectic duality. The Schrödinger equation in its standard complex form describes the unitary evolution of a quantum system, while its real form describes the symplectic evolution of a classical mechanical system.
This quantum-symplectic duality can be leveraged to enhance the capabilities of quantum and symplectic computers. In this symplectic approach, the role of a quantum bit (qubit) is taken by a symplectic bit (symbit).

}
\maketitle

\section{Introduction}
Recent years have seen significant progress in the field of quantum computer technology,  see, for example, \cite{nature, VolOhya}.
\\

In this paper, we propose the  concept of symplectic computers, which have the potential to be more powerful than quantum computers. Unlike quantum computing, which consists of a sequence of unitary transformations (gates) and projectors (measurements), symplectic computation involves a sequence of symplectic transformations and measurements. The main mathematical idea of  the symplectic computers is based on the quantum-symplectic duality.  The Schrödinger equation in its standard complex form describes the unitary evolution of a quantum system, while its real form describes the symplectic evolution of a classical mechanical system.
This quantum-symplectic duality can be leveraged to enhance the capabilities of  symplectic computers. In this symplectic approach, the role of a quantum bit (qubit) is taken by a symplectic bit (symbit).
\\

Above mentioned quantum-symplectic duality can be explained as follows.
We consider the final dimensional Hilbert space ${\mathbb C}^N$.
The Schrodinger equation in its standard complex form reads
\be
\label{Sch}
 i \dot \psi=H\,\psi.\ee
 Here $\psi$ is a complex valued wave function, the dot means the time derivative and $H$ is an hermitian matrix. This matrix  $H$ can be represented in the form
 \begin{equation}\label{HKL}
     H = K+i L,
 \end{equation}
where $K$ is a symmetric real matrix and $L$ is a skew real matrix. Now we rewrite  equation \eqref{Sch} in the real form by using the representation of the wave function 
 \be
\psi=q+i p,
\ee
where $p$ and $q$ are the real and imaginary parts of $\psi$. Then the  Schrodinger equation \eqref{Sch} takes the form
\bea\label{dot-q}
\dot q &=&\,\, \,Kp+Lq,\\\label{dot-p}
\dot p &=&- Kq+Lp.\eea
The Schrodinger equation in the real form \eqref{dot-q}, \eqref{dot-p} describes a symplectic evolution of the corresponding mechanical system \cite{IV,Volovich:2019rxl}. 
\\

This fact can be called the quantum-symplectic duality.  The idea of such duality comes back to the wave-particle duality suggested by Einstein \cite{AE} and de Broglie \cite{deBroglie:1924ldk}. From  the modern point of view   wave-particle duality is the concept in quantum mechanics that quantum entities exhibit particle or wave properties according to the experimental circumstances \cite{RF,PMD}. \\

The unitary group is a subgroup of a symplectic group and we have the relation 
\be\label{1.6}
    U(N)=Sp(2N, {\mathbb R}) \cap {\mathbb O}(2N, {\mathbb R}).
\ee
Therefore, any quantum computation is a symplectic computation. However, not every symplectic computation is quantum. For this reason, symplectic computers are, in principle, generally more powerful than quantum ones.
\\

The paper is organized as follows. In Section \ref{QSD} we discuss the 
quantum-symplectic duality. In Subsection \ref{subs:SESM} the Schrodinger equation is presented  as symplectic mechanics. In  Subsection \ref{subs:USE} unitary and symplectic evolutions  are discussed. In Subsection \ref {subs:QS} qubits and symbits are compared. In  Section \ref {sec:US} unitary and symplectic gates are considered.
Possible generalizations of the proposed approach are mentioned  in Section \ref{sec:CD}.


\section{Quantum-symplectic duality}\label{QSD}

\subsection{Schrodinger equation and symplectic mechanics}\label{subs:SESM}

The Schrodinger equation actually describes the classical symplectic  Hamiltonian system.
We consider the  finite dimensional Hilbert space $\mathcal{H}=\mathbb{C^N}$. The Schrodinger equation reads
\be \label{Schreq}
    i \dot {\psi_a} = H_{ab} \psi_b
\ee
where $a,b=1...N$, $\psi_a \in \mathbb{C} $, $H_{ab}=\overline H_{ba}$ is an Hermitian matrix. 
  We assume the summation over the repeating indices.  

We rewrite this equation by using real quantities and set 
\be \label{psia}\psi_a = q_a + ip_a,
\ee 
where $q_a, p_a$ are real. 
We also write 
\be\label{Hab}H_{ab} = K_{ab}+i L_{ab},
\ee
where $K_{ab}=K_{ba}$, $L_{ab} = - L_{ba}$ are  matrices with real entries.
Then the Schrodinger equation \eqref{Schreq} takes the form
\begin{equation}
    \dot{q_a} = K_{ab} p_b + L_{ab} q_b
\end{equation}
\begin{equation}
    \dot{p_a} = -K_{ab} q_b + L_{ab} p_b
\end{equation}
These equations are classical symplectic  Hamiltonian equations 
\bea 
\dot q_a=\frac{\partial H_{sym}}{\partial p_a},\qquad\dot p_a=-\frac{\partial H_{sym}}{\partial q_a},
\eea
with the Hamiltonian 
\be
    H_{sym}= \frac{1}{2} (p_a K_{ab}\, p_b+ q_a K_{ab}\, q_b) + p_a L_{ab}\,q_b 
\ee
\\

\textbf{Remark 1.} If the Hamiltonian equation in the Schrodinger equations
depends on time
\be \label{Schreq-t}
    i \dot {\psi_a} = H_{ab}(t) \psi_b
\ee
with
\be
\label{Habt}
H_{ab}(t) = K_{ab}(t)+i L_{ab}(t),
\ee
then we get the following equation

\begin{equation}
    \dot{q_a} = K_{ab}(t) p_b + L_{ab}(t) q_b
\end{equation}
\begin{equation}
    \dot{p_a} = -K_{ab}(t) q_b + L_{ab}(t) p_b
\end{equation}
 \be
    H_{sym}(t)= \frac{1}{2} (p_a K_{ab}(t)\, p_b+ q_a K_{ab}(t)\, q_b) + p_a L_{ab}\,q_b. 
\ee
Note in particular that it seems the problem of quantum control can be reduced the problem of classical control theory by using quantum-symplectic duality, see \cite{Pechen} for a discussion of the quantum control theory.

\subsection{Unitary and symplectic evolution} \label{subs:USE}
In this subsection we discuss the relation between representations of solutions of the Schrodinger/Hamiltonian equations in the complex unitary form and the real symplectic form.

The solution of the N-component  Schrodinger equation 
\begin{equation}
    \psi (t) = U_t \,\psi(0), \qquad \psi(t) = \left(\psi_1(t),...\psi_N(t)\right),
\end{equation}
can be rewritten in term of real N-component vectors
\be
\psi(t)=q(t)+i p(t).
\ee
Representing the unitary matrix $U_t$ in the form 
\begin{equation}
    U_t = X_t + iY_t,
\end{equation}
where $X_t$ and $Y_t$ are real matrices, 
we have
\bea
\psi(t)=X_tq(0)-Y_tp(0)+i(Y_tq(0)+X_tp(0)),
\eea
i.e.
\bea
q(t)&=&X_tq(0)-Y_tp(0)\\
p(t)&=&Y_tq(0)+X_tp(0),\eea
These relations  can be represented in the form  
\be\label{2.7}
 \begin{pmatrix}
 q(t)\\p(t)    
 \end{pmatrix}   
= S_t \begin{pmatrix}
 q(0)\\p(0)    
 \end{pmatrix},
\end{equation}
where  $S_t$ given by 
\begin{equation}
S_t = \left(\begin{array}{ll}
\,\,\,X_t & Y_t \\
-Y_t & X_t
\end{array}\right).
\end{equation}

The unitarity of matrix $U_t$,
\be
U_t ^*\,U_t=(X_t^T - iY_t^T)
(X_t + iY_t)=X_t^TX_t+Y_t^TY_t
-i(Y_t^TX_t-X_t^TY_t)=I
\ee
means that 
\bea\label{2.16}
X_t^TX_t+Y_t^TY_t&=&I,\\
\label{2.17}
Y_t^TX_t-X_t^TY_t&=&0\eea
These relation imply that  $S_t$ is a symplectic matrix, i.e.   satisfies the relation 
\be\label{Smat}
S^T_t  J S_t = J,
\ee
here $S^T$ means the transposed  matrix and 
\be
J=\left(\begin{array}{ll}
\,\,\,0 & I \\-I& 0
\end{array}\right)
\ee

Indeed, 
\bea\nn
S^T_t  J S_t&=&
\left(\begin{array}{ll}
\,\,\,X_t^T Y_t-Y_t^TX_t&\,\, Y_t^T Y_t+X_t^T X_t\\-Y_t^TY_t-X_t^TX_t&\,  -Y_t^TX_t+X_t^TY_t
\end{array}\right)\eea
and taking into account \eqref{2.16} and \eqref{2.17} we get 
\bea
S^T_t  J S_t&=&
\left(\begin{array}{ll}
\,\,\,0 & I \\-I& 0
\end{array}\right),\eea
i.e. equation \eqref{Smat}.
\subsection{Qubits and Symbits}\label{subs:QS}

In quantum computing, the fundamental unit of information is the qubit. The qubit is a two-level quantum system, mathematically represented by the complex Hilbert space $\mathbb{C}^2$. A general state of a qubit can be written as:
\be\label{s-q}
|\psi\rangle = \alpha |0\rangle + \beta |1\rangle, \quad \alpha, \beta \in \mathbb{C}, \quad |\alpha|^2 + |\beta|^2 = 1,
\ee
where $|0\rangle$ and $|1\rangle$ form an orthonormal basis for $\mathbb{C}^2$. The coefficients $\alpha$ and $\beta$ are complex numbers such that the total probability (norm) is equal to one.
\\
 
In the symplectic computing framework, the analogous concept to a qubit is a symplectic bit (symbit). A symbit is represented by a two-dimensional real vector space $\mathbb{R}^2$. A general state of a symbit can be written as:
\be 
\label{s-s}
|\phi) = \alpha |0) + \beta |1), \quad \alpha, \beta \in \mathbb{R}, \quad \alpha^2 + \beta^2 = 1,
\ee
where $|0)$ and $|1)$ form an orthonormal basis for $\mathbb{R}^2$. 
The coefficients $\alpha$ and $\beta$ here are real numbers such that the total probability  is equal to one.
We can interpret $\mathbb{R}^2$ as the phase plane for a dynamical system with coordinates $p$ and $q$.
\\

One can consider tensor products of $N$ sbits:
\be
\mathbb{R}^2 \otimes \mathbb{R}^2 \otimes \cdots \otimes \mathbb{R}^2 = \mathbb{R}^{2^N}.
\ee
This is equivalent to \(\mathbb{C}^N\). On the other hand, the tensor product of \( N \) qubits is \(\mathbb{C}^{2^N}\). Therefore, by using tensor products of sbits, one can obtain more general spaces than those obtained from tensor products of qubits.
\\

Operations on symbits are performed using symplectic transformations, which are elements of the symplectic group $Sp(2, \mathbb{R})$. For instance, the simplest symplectic transformation in $\mathbb{R}^2$ can be represented by a $2 \times 2$ matrix $S$ that preserves the symplectic form:
\[
S^T J S = J, \quad \text{where} \quad J = \begin{pmatrix}
0 & 1 \\
-1 & 0
\end{pmatrix}.
\]
An example of a symplectic transformation is the rotation matrix:
\[
R(\theta) = \begin{pmatrix}
\cos\theta & \sin\theta \\
-\sin\theta & \cos\theta
\end{pmatrix},
\]
which preserves the symplectic form and hence is a member of $Sp(2, \mathbb{R})$.
\\

In summary, while qubits are the basic units of quantum information in $\mathbb{C}^2$ and are manipulated using unitary transformations, symbits are the basic units of symplectic information in $\mathbb{R}^2$ and are manipulated using symplectic transformations. This duality provides a bridge between quantum and symplectic computing, offering a new perspective on computational processes.

\section{Unitary and symplectic gates}\label{sec:US}
\subsection{Unitary group $U(N)$ and symplectic group $Sp(2N, {\mathbb R})$}
Let us show  that unitary group $U(N)$ is a subgroup of symplectic group $Sp(2N, {\mathbb R})$. Let us remind that a symplectic matrix $S$ of symplectic group $Sp(2N, \mathcal{R})$ satisfies the relation 
\begin{equation}
    S^T J S = J
\end{equation}
where 
 \be J = \left( \begin{array}{ccc} 0~~~I\\-I~~~0 \end{array}\right).\ee 

 Let $S$ is a $2 N\times 2N$ matrix in the form
\begin{equation*}
S=\left(\begin{array}{ll}
A & B \\
C & D
\end{array}\right)
\end{equation*}
where $A, B, C, D$ are $N \times N$ matrices. 
The conditions on $A,B,C$ and $D$ are

\begin{itemize}
    \item $A^{\mathrm{T}} C, B^{\mathrm{T}} D$ symmetric, and $A^{\mathrm{T}} D-C^{\mathrm{T}} B=I$
    \item $A B^{\mathrm{T}}$, $C D^{\mathrm{T}}$ symmetric, and $A D^{\mathrm{T}}-B C^{\mathrm{T}}=I$
\end{itemize}

Let us show that there is a canonical mapping $\gamma$ of the unitary group $U(N)$ to $Sp(2N, {\mathbb R})$. Let $V \in U(N)$, we present $V$ in the form 
\begin{equation}
    V=X+iY, ~~~ \text{where } X,Y \text{ are } N\times N \text{matrices with real entries}
\end{equation}
We define $\gamma$ by the following formula 
\begin{equation} \label{gamma}
    \gamma(V) = \gamma(X+iY)= 
    \left(\begin{array}{ll}
X & Y \\
-Y & X
\end{array}\right)
\end{equation}
One can check that the unitarity conditions $VV^* = V^*V =1$ lead to the conditions for the symplectic matrices.
\\

\textbf{Remark 2.} One can also check  that there is  a  relation $\gamma (V_1) \gamma (V_2)=\gamma(V_1V_2)$.
\\

\textbf{Remark 3.} The Schrodinger equation in the form 
\be
i \dot \psi=H\,\psi,\ee
where \be H=-\Delta+V(x)
\ee
$\Delta$ is the Laplace operator in $\mathbb R$, in  the real formulation looks as follows
\be
\dot q=H p,\qquad \dot p=-H q.\ee
Spectrum and scattering theory for in this approach are considered in \cite{IVVG}

\subsection{NOT and $\gamma$(NOT) gates}

Quantum computation is a sequence of unitary matrices (gates) of the simple form. We put into correspondence into such unitary gate a corresponding symplectic gate using the $\gamma$ defined by \eqref{gamma}.

See the simplest gate is NOT. It is a unitary matrix $NOT = \left(\begin{array}{ll}
0 & 1 \\
-1 & 0
\end{array}\right)$ defined on the qubit $\mathbb{C}^2$. By using the mapping $\gamma$ we define the corresponding symplectic gate 
\begin{equation}
    \gamma(NOT) = 
    \left(\begin{array}{llll}
0 & 1 & 0 & 0\\
-1 & 0 & 0 & 0 \\
0 & 0 & 0 & 1\\
0 & 0 & -1 & 0
\end{array}\right)
\end{equation}
One can construct the symplectic analog of the controlling CNOT gate.

\section{Conclusion and Discussion}\label{sec:CD}

In this paper, we have introduced the concept of symplectic computers and proposed that they have the potential to offer computational advantages over quantum computers. By utilizing the quantum-symplectic duality, we demonstrated that the Schrödinger equation, when expressed in real terms, corresponds to a classical symplectic mechanical system. This allows for the description of quantum systems in the framework of symplectic mechanics, with symplectic bits (symbits) serving as analogs to quantum bits (qubits).
\\

We have remind how the classical symplectic Hamiltonian equations can be derived from the Schrodinger equation, enabling the use of symplectic transformations and projectors in symplectic computation, similar to unitary transformations and measurements in quantum computation. Our examination of the relationship between unitary and symplectic evolutions revealed that unitary matrices can be mapped to symplectic matrices, ensuring that quantum gates have corresponding symplectic counterparts. This suggests that symplectic computing can replicate the capabilities of quantum computing while potentially providing additional benefits.
Additionally, we discussed the implications of symplectic mechanics for quantum computation, suggesting that quantum control problems might be addressed using classical control theory through quantum-symplectic duality. This could lead to new insights and methods in both quantum and symplectic computing.
\\

 Let us note that our discussion has focused on linear symplectic computation. Future research could explore non-linear symplectic transformations and the application of dynamics on symplectic manifolds. Note that the complexity of some manifolds, including black holes, has already been discussed \cite{Susskind:2014rva}. It would also be interesting to explore computational devices based on non-quadratic Hamiltonian systems, such as the nonlinear Schrodinger equation. This could provide a connection with approaches to quantum theory hidden in classical probability \cite{LA}.
Here, we discussed the finite-dimensional Hilbert space. We expect that our considerations can be extended to the infinite-dimensional Hilbert space in line with the considerations in the papers \cite{IV,Volovich:2019rxl,SakVol-JSA}.
\\

In this paper, we have discussed quantum and symplectic computers, which consist of products of unitary or symplectic matrices. It seems natural to extend the concept of computations to include sequences of elements from arbitrary Lie groups. From this perspective, unitary representations of Lie groups would be natural extensions of quantum computers. As a further generalization, the $q$-deformations of Lie groups could be considered. Finally, one can study computations on various number fields, including $p$-adic, see 
\cite{IV-padic,30years,Anashin}.
\\

In summary, the introduction of symplectic computers presents an interesting development in the field of computation. By further exploring the theoretical foundations and practical implementations of symplectic computing, we may discover new computational techniques and applications that enhance both quantum and classical computing paradigms.

\section{Acknowledgements}

I am very grateful to I. Ya. Aref’eva, V. Zh. Sakbaev, and D. Stepanenko for fruitful
discussions. 
This work is supported
by the Russian Science Foundation (project  24-11-00039, V.A. Steklov Mathematical
Institute)

\end{document}